\documentstyle[aps,prl,multicol]{revtex}
\begin{document}
\draft
\title{On the Turbulent Dynamics of Polymer Solutions}
\author{E.~Balkovsky$^{a,b}$, A.~Fouxon$^a$, and V.~Lebedev$^{a,c}$}
\address{$^a$ Physics Department, Weizmann Institute of Science,
Rehovot 76100, Israel \\ $^b$ School of Mathematics, Institute for
Advanced Study, Einstein Drive, Princeton, NJ 08540, USA \\ $^c$ Landau
Inst. for Theor. Physics, Moscow, Kosygina 2, 117940, Russia}
\date{\today}

\maketitle

\begin{abstract}

We study properties of dilute polymer solutions which are known to
depend strongly on polymer elongation. The probability density
function (PDF) of polymer end-to-end extensions $R$ in turbulent flows
is examined. We demonstrate that if the value of the Lyapunov exponent
$\lambda$ is smaller than the inverse molecular relaxation time
$1/\tau$ then the PDF has a strong peak at the equilibrium size $R_0$
and a power tail at $R\gg R_0$. This confirms and extends the results
of \cite{Lumley72}. There is no essential influence of polymers on the
flow in the regime $\lambda\tau<1$. At $\lambda>1/\tau$ the majority
of molecules is stretched to the linear size $R_{\rm op}\gg R_0$. The
value of $R_{\rm op}$ can be much smaller than the maximal length of
the molecules because of back reaction of the polymers on the flow,
which suppresses velocity gradients thus preventing the polymers from
maximal possible stretching.

\end{abstract}

\pacs{PACS numbers 83.50.Ws, 61.25.Hq, 47.27.-i, 05.40.-a}

\begin{multicols}{2}

Dynamics of dilute polymer solutions is an important subject both from
theoretical and practical points of view. Possible applications rely
mainly on the fact that low concentrations of polymer molecules can
lead to substantial changes in hydrodynamics. The most striking effect
related to polymers is probably the so-called drag reduction in
turbulent flows. A consistent explanation of this effect is a
long-standing question \cite{BCA}. One believes that the drag
reduction is related to the effective increase of the viscosity due to
the presence of polymers \cite{Lumley73}. Here we address some aspects
of this phenomenon.

An important underlying property of polymers is their flexibility
\cite{Lumley72,Lumley73,Lumley69,Hinch77}. At equilibrium, a polymer
molecule coils up into a spongy ball with a radius $R_0$. For dilute
solutions with concentrations $n$ satisfying $nR_0^3\ll 1$, the
influence of equilibrium size molecules on hydrodynamic properties can
be neglected. When placed in a flow, the molecule is deformed into an
elongated structure of ellipsoidal form which can be characterized by
its end-to-end extension $R$. Since the number $N$ of monomers in a
long-chain polymer molecule is large, $R$ can be much larger than
$R_0$. It explains why minute amounts of polymers can produce an
essential effect. It was shown in \cite{Lumley72} that in sufficiently
intensive flows polymer molecules get strongly extended due to
stretching.  This is the key mechanism providing an essential back
reaction of the polymer molecules on the flow.

Here we consider turbulent dynamics of polymer solutions. We assume
that $R$ is always much smaller than the viscous length of the
turbulent flow, $r_v$. Therefore, molecules can be treated as immersed
into a spatially smooth external velocity field \cite{Lumley69}. In
this case the dynamics of polymer stretching is determined only by the
gradients of the velocity. Since the gradients in turbulent flows are
correlated at the viscous length, all the molecules inside regions
with size of the order of $r_v$ are subject to the same gradient, and
therefore are stretched coherently. As long as one can neglect the
hydrodynamic interactions between molecules, the problem is reduced to
dynamics of a single molecule.

We investigate the behavior of polymer molecules with the extensions
$R$ satisfying $R_0\ll R\ll R_{\rm max}$, where $R_{\rm max}$ is the
maximal size of the polymer. Random walk arguments show that the
entropy of such molecules is quadratic in $R$ in this interval, which
leads to Hooke's law (see e.g. Ref. \cite{Kit}). That is why one can
expect a linear dynamics of the molecules. Even though hydrodynamic
interactions of monomers make polymer's dynamics inherently nonlinear,
the interactions can be neglected for elongated molecules. This
expectation is confirmed by recent experiments with DNA molecules
\cite{Chu1} where an exponential relaxation of a single molecule was
observed.  Numerics and theoretical arguments presented in
Ref. \cite{99HQ} also show the linear character of the molecule
dynamics for $R_0\ll R\ll R_{\rm max}$. In experiments
\cite{Chu1} a number of the molecule eigen modes has been seen. We
will take into account only the mode which has the largest relaxation
time $\tau$, because the other modes are harder to get excited in
turbulent flows.

A starting point of our theory is the dynamic equation for the vector
${\bbox R}$ which can be defined, say, via the inertia tensor (per
mass of a monomer) $R_\alpha R_\beta$ of the elongated molecule. Then
${\bbox R}$ determines the orientation and the largest size of the
molecule. We assume the following dynamic equation for the vector
(cf. Refs. \cite{Lumley73,Hinch77})
\begin{eqnarray}
\frac{{d}}{{d}t}R_\alpha=R_\beta
\nabla_\beta v_\alpha
-\frac{R_\alpha}{\tau}\,,
\label{basic} \end{eqnarray}
where $\tau$ is the relaxation time. The velocity gradient must be
taken at the molecule position. The role of non-linearity in the
extended equation for ${\bbox R}$ (and in the system of equations for
$N$ coupled beads) is examined in Ref. \cite{misha}. For our purposes
this non-linearity as well as the thermal noise is irrelevant (see the
discussion below).

To get rid of inessential degrees of freedom responsible for the
orientation of the molecule we write ${\bbox R}=R{\bbox n}$ passing to
the absolute value $R$ of the vector ${\bbox R}$. Then we obtain from
Eq. (\ref{basic}) (cf. \cite{95CFKL})
\begin{eqnarray} &&
\frac{{d}\rho}{{d}t}=\zeta-\frac{1}{\tau} \,,
\qquad \frac{{d}n_\alpha}{{d}t}=n_\beta
\nabla_\beta v_\alpha -\zeta n_\alpha \,,\qquad 
\label{basic1} \\ &&
R=R_0\exp(\rho) \,, \qquad
\zeta=n_\alpha n_\beta\nabla_\beta v_\alpha \,.
\label{basic2} \end{eqnarray}
We see that the evolution of $\rho$ is determined by the scalar
function $\zeta$ which is a functional of the velocity field.

For turbulent flows where the velocity randomly varies in time one
should use a statistical approach. A natural first step is to take the
polymers being passively embedded into the fluid, disregarding their
back reaction on the flow. We will demonstrate that there is a wide
region of applicability of this approximation. Neglecting the back
reaction we can treat the velocity dynamics as independent of
polymers. Then $\zeta$, defined by Eq. (\ref{basic2}) is 
independent of $\rho$. We will not specify the velocity
statistics. Irrespective of its character one can use the large
deviation theory (see e.g. Ref. \cite{BF99} devoted to different
aspects of Lagrangian dynamics in turbulent flows). The scheme
presented below is valid for any random flow. Integrating
Eq. (\ref{basic1}) we get
\begin{eqnarray}&&
\rho(t)=\rho_0+z-\frac{t}{\tau}\,, \qquad
z=\int_0^t \!{d}t'\,\zeta(t') \,,
\label{sol1} \end{eqnarray}
where $\rho_0$ is the value of $\rho$ at $t=0$. One should keep in
mind that the expression (\ref{sol1}) for $\rho$ is correct if one can
neglect the presence of the boundaries $R_0$ and $R_{\rm max}$ where
Eq. (\ref{basic}) is violated.

The integral $z$ in Eq. (\ref{sol1}) possesses some universal
properties for times much larger than the correlation time
$\tau_\zeta$ of the random process $\zeta$. For turbulent flows
$\tau_\zeta$ can be estimated as the characteristic time of the
Lagrangian motion on the viscous scale, which coincides with the
characteristic inverse strain on this scale. For $t\gg\tau_\zeta$ the
variable $z$ can be considered as a sum of a large number of
independent variables. Then in order to establish the statistics of
$z$ for fluctuations near its mean value one can use the central limit
theorem. If we are interested in large deviations from the mean, a
more general formulation is needed (see
e.g. \cite{Frish,Lanford}). Namely, the PDF of $z$ can be written as
the homogeneous function
\begin{eqnarray} &&
{\cal G}(t,z)\approx \frac{1}{\sqrt{2\pi \Delta t}}
\exp\left[-tS\left(\frac{z-\lambda t}{t}\right) \right]\,,
\label{entr} \\ &&
\lambda=\left\langle\zeta\right\rangle\,,\quad
\Delta=\int {d}t'\,\left(\left\langle\zeta(t)
\zeta(t')\right\rangle-\lambda^2\right)\,.
\label{lambda} \end{eqnarray}
``The entropy density'' $S$ is a functional of the velocity
statistics. It is impossible to calculate $S$ without knowing the
statistics explicitly. Fortunately, only general properties of $S$
(such as positivity and convexity) are needed for us. The central
limit theorem is reproduced by Eq. (\ref{entr}) if to consider a
vicinity of the entropy maximum where $S(x)\approx x^2/(2\Delta)$. The
constant $\lambda$ defined in (\ref{lambda}) is the principal Lyapunov
exponent of the turbulent flow, which is the average logarithmic rate
of growth of the distance between two initially close Lagrangian
trajectories.

As follows from Eq. (\ref{sol1}), ${\cal G}(t,z)$ determines the
conditional probability that $\rho(t)$ has the value $\rho_0+z-t/\tau$
provided $\rho(0)=\rho_0$. Therefore one can write the equation
\begin{eqnarray}
{\cal P}(t,\rho)=\int{d}\rho_0\,
{\cal G}(t,\rho-\rho_0+t/\tau){\cal P}(0,\rho_0)
\label{stat} \end{eqnarray}
for the PDF ${\cal P}(t,\rho)$. In the stationary case ${\cal P}$ is
$t$-independent and Eq.  (\ref{stat}) can be treated as a relation
determining the PDF. Writing ${\cal P}$ as the Laplace integral ${\cal
P}(\rho)=\int{d}\alpha\,\exp(-\alpha\rho)\tilde{\cal P}(\alpha)$ we
observe that the convolution in Eq. (\ref{stat}) becomes a product and
the equation can be easily resolved. The value of $\tilde{\cal
P}(\alpha)$ is non-zero if
\begin{eqnarray}
\int\frac{{d}x}{\sqrt{2\pi \Delta t}} 
\exp\left[\alpha x
-tS\left(\frac{x}{t}
+\frac{1}{\tau}-\lambda\right) \right]=1 \,.
\label{tail1} \end{eqnarray}
Apart from the trivial solution $\alpha=0$ this equation defines
$\alpha$ uniquely.

Since $t\gg\tau_\zeta$, one can use the saddle-point approximation
in calculating the integral (\ref{tail1}). It gives the condition
\begin{eqnarray}
\alpha=S'(\beta+1/\tau-\lambda) \,,
\label{alpha} \end{eqnarray}
where $\beta$ is the saddle-point value of the ratio
$(\rho-\rho_0)/t$. Equating the integral in the left-hand side of
Eq. (\ref{tail1}) (calculated in the same approximation) to unity we
get the equation for $\beta$
\begin{equation}
S\left(\beta-\lambda+\frac{1}{\tau}\right)-\beta
S'\left(\beta-\lambda+\frac{1}{\tau}\right)=0\,.
\label{beta} \end{equation}
It is important that $\beta$ is independent of $t$ and $\rho$. Solving
Eq. (\ref{beta}) and substituting the result into Eq. (\ref{alpha}) we
find the exponent $\alpha$. The trivial solution
$\beta=\lambda-1/\tau$ of Eq. (\ref{beta}), corresponding to
$\alpha=0$, should be discarded.

We conclude that a single component $\tilde{\cal P}(\alpha)$ is
non-zero and therefore ${\cal P}(\rho)\propto\exp(-\alpha\rho)$.
Recalculating this distribution of $\rho$ into that of $R$ we obtain
the power tail of the PDF of the molecule size $R$
\begin{equation}
{\cal P}(R)\sim R_0^{\alpha}R^{-\alpha-1}\,.
\label{pdf}\end{equation}
For positive $\alpha$ the normalization integral $\int{d}R\,{\cal
P}(R)$ is determined by small $R$, which means that the majority of
molecules has nearly equilibrium size. On the contrary, the
normalization integral diverges at large $R$ if $\alpha<0$. Then the
majority of molecules is strongly stretched.

Another way to obtain the result (\ref{pdf}) is to consider the
typical fluctuation making the largest contribution into the tail of
the PDF. Starting from a nearly equilibrium shape, that is from
$\rho_0\sim 1$, the velocity stretches the molecule up to $\rho\gg
1$. The contribution of fluctuations with stretching period $t$ is
equal to ${\cal G}(t,\rho+t/\tau)$. It has a sharp maximum at time
$t_*$ determined from $d{\cal G}(t,\rho+t/\tau)/dt=0$. This condition
gives $t_*=\rho/\beta$. The probability density is thus dominated by
fluctuations with stretching period $t_*$.  It is proportional to
${\cal G}(t_*,\rho+t_*/\tau)$ which reproduces Eq. (\ref{pdf}) with
$\alpha$ given by Eq. (\ref{alpha}). Note that the characteristic
value of the velocity gradient for the relevant fluctuations is given
by $\zeta\sim\rho/t_*+1/\tau$ and is of the order $1/\tau$.

Let us establish the dependence of $\alpha$ on the control parameter,
which is the strength of velocity fluctuations at the viscous length
measured by the Lyapunov exponent $\lambda$. As $\lambda$ tends to
zero, the function $S(x)$ contracts to $x=0$ and therefore $\alpha$
tends to infinity, which implies strong suppression of the tail. It is
quite natural since in a weak flow the molecules are only weakly
stretched. Note that even for intense flows the Lyapunov exponent
$\lambda$ is suppressed in the regions where the rotation rate
dominates the strain rate. As ${\lambda}$ increases, the exponent
$\alpha$ decreases and at a certain level of fluctuations approaches
zero. If $\lambda$ is close to $1/\tau$ then one can use the quadratic
approximation for $S$ which leads to the law
\begin{equation}
\beta=\frac{1}{\tau}-\lambda \,, \qquad
\alpha=\frac{2}{\Delta}\left(
\frac{1}{\tau}-\lambda\right) \,.
\label{crit} \end{equation}
We see that $\alpha$ changes its sign at $\lambda=1/\tau$. Thus the
majority of molecules becomes stretched when $\lambda>1/\tau$. This
can be interpreted as the criterion for the coil-stretch transition in
turbulent flows discussed in \cite{Lumley72,Lumley73,gennes}.

We can use Eq. (\ref{stat}) only if $\rho$ and $\rho_0$ belong to the
asymptotic region between zero and $\rho_{\rm max}$, where
Eq. (\ref{basic}) is valid. The saddle-point approximation used above
gives $\rho_0=\rho-\beta t$. Thus the above scheme works only if
$t<t_*=\rho/\beta$ (here we assume $\beta>0$, i.e. $\alpha>0$). Then
the polymer molecules spend most of the time fluctuating near the
equilibrium shape, occasionally getting stretched by strain
fluctuations which overcome the elastic reaction. The fluctuations
leading to a given $R$ have the duration $t_*\approx
\rho/\beta$. Since $\beta$ tends to zero when $\lambda\to1/\tau$ one
should observe a critical behavior $t_*\propto|\lambda-1/\tau|^{-1}$
in accordance with Eq. (\ref{crit}). We see that in the vicinity of
$\lambda=1/\tau$ the time $t_*$ is much larger than $\tau_\zeta$ which
justifies our scheme. Similar considerations are valid for $\alpha<0$.

One can generalize the scheme taking into account a number of
molecular eigen modes. Since the critical value of $\lambda$ is
determined by the inverse relaxation time, then in the vicinity of the
critical value corresponding to the principal mode, the other modes
are at most weakly excited. However, they can be important at larger
$\lambda$.

The rest of the paper is devoted to the discussion of the back
influence of the polymers on the flow. A consistent investigation
should be based on the complete system of equations coupling
turbulence with polymers. One of these equations is the modified
Navier-Stokes equation
\begin{eqnarray}&&
(\partial_t +{\bbox v}\nabla)v_\alpha
=-\nabla_\alpha p+\nu\nabla^2 v_\alpha
+\nabla_\beta\Pi_{\alpha\beta}\,,
\label{navier} \end{eqnarray}
where $\Pi_{\alpha\beta}$ is the polymer contribution to the stress
tensor. Equation (\ref{navier}) should be supplemented with the
equation describing dynamics of $\Pi_{\alpha\beta}$. In the considered
case $\Pi_{\alpha\beta}$ can be defined as a sum of stresses of
polymer molecules in a volume divided by the mass of the fluid inside
the volume \cite{LL6}. We are interested in the situation when the
molecules are strongly elongated. Then due to Hooke's law the stress
of such molecule is proportional to $R_\alpha R_\beta$. Next, taking
the volume smaller than the viscous length we deal with coherently
elongated molecules. Therefore the stress tensor can be written as
\begin{eqnarray}
\Pi_{\alpha\beta}=\Pi_0 \exp(2\rho)n_\alpha n_\beta \,,
\label{stress} \end{eqnarray}
where $n_\alpha$ is a unit vector, $\Pi_0 \exp(2\rho)$ is the
principal eigenvalue of $\Pi_{\alpha\beta}$ and the elongated
molecules correspond to $\rho>0$. Then from Eq. (\ref{basic}) we get
the same Eqs. (\ref{basic1}) for $\rho$ and ${\bbox n}$, where
${d}/{{d} t}$ should be understood as the material derivative
$\partial_t+{\bbox v}\nabla$. Thus the velocity ${\bbox v}$ is now
coupled to $\rho$ and ${\bbox n}$ via
Eqs. (\ref{navier},\ref{stress}). Note that the constant $\Pi_0$ in
Eq. (\ref{stress}) is proportional to the concentration of the polymer
molecules.

Let us consider the PDF of $R$ not assuming that the flow is
unperturbed by the polymers. We start from the case
$\lambda\tau<1$. One recovers Eq. (\ref{pdf}) if the back influence is
small, i.e. $\Pi\ll\nu\nabla v$ for the relevant fluctuations
characterized by $\nabla v\sim 1/\tau$. Since $\Pi\propto R^2$, the
polymer contribution in the stress tensor grows with $R$ and the
inequality $\Pi\ll\nu\nabla v$ is violated for the molecules with
$R\gtrsim R_{\rm back}$. The value of $R_{\rm back}$ can be found from
the estimate $\nu/\tau\sim\Pi_0 R_{\rm back}^2/R_0^2$. For $R\gtrsim
R_{\rm back}$ the back reaction switches on and suppresses the
velocity fluctuation. Hence, the probability of fluctuations producing
$R>R_{\rm back}$ is small and hence at $R\gtrsim R_{\rm back}$ the PDF
decays much faster than prescribed by Eq. (\ref{pdf}).

Now we study the case $\lambda>1/\tau$. For $R\ll R_{\rm back}$ the
polymer stress tensor is small and as explained above velocity is
decoupled from the elastic degrees of freedom. Since stretching is
stronger than the elastic force, $R$ grows for any typical velocity
realization. On the other hand, at $R\gtrsim R_{\rm back}$ the polymer
stress influences the velocity, suppressing it strongly for
sufficiently large $R$. This leads to a decrease in $R$. Therefore,
the majority of molecules has sizes near an optimal size $R_{\rm
op}>R_{\rm back}$. The PDF is an increasing function of $R$ at
$R<R_{\rm op}$ and decays fast at $R> R_{\rm op}$. In this state
velocity gradients can be estimated as $1/\tau$ \cite{BFLprep}. This
can be proven e.g. by averaging Eq. (\ref{basic1}). It means that the
Lyapunov exponent of the solution is smaller than that of the solvent
at the same energy input. The energy dissipation is related mainly to
the polymer stress tensor and hence $R_{\rm op}$ grows as the input of
energy increases
\cite{BFLprep}. The effective viscosity defined as the proportionality
coefficient between $\Pi$ and $\nabla v$, also grows. Note that there
exists an interval where $R_{\rm op}\ll R_{\rm max}$. This contradicts
the widely accepted view that at some level of turbulent fluctuations
there is a sharp transition between the state where most of the
molecules have $R\sim R_0$ and the state where all the molecules are
stretched up to $R_{\rm max}$.

We conclude that for $\alpha>0$ (i.e. $\lambda<1/\tau$) the end-to-end
extensions of the majority of molecules are of the order of the
equilibrium size $R_0$, and there is no essential contribution to the
stress tensor. For $\alpha<0$ (i.e. $\lambda>1/\tau$) extensions of
most of the molecules are of the order of $R_{\rm op}\gg R_0$. Then
the polymer stress tensor $\Pi$ is estimated as $\Pi_0R^2_{\rm op}/
R^2_0$. Its value can be much larger than the viscous contribution
$\nu\lambda$ \cite{BFLprep}.

The above analysis implies that $R_{\rm back}\ll R_{\rm max}$ since at
$R\sim R_{\rm max}$ one must consider non-linear corrections to
Hooke's law and hence to Eq. (\ref{basic}). The condition $R_{\rm
back}\ll R_{\rm max}$ is realized for sufficiently high concentrations
of polymers. Then the fluid displays non-Newtonian properties. When
most of the molecules are stretched up to $R_{\rm max}$ but the back
reaction is not switched on, one has a Newtonian fluid whose
properties do not differ significantly from the properties of the
solvent. This is the case for very dilute solutions where
$\Pi\ll\nu\nabla v$.

Since a turbulent flow is multiscale, the real picture is more
complicated. We have shown that the main characteristics of the flow
that determines the behavior of polymer molecules is the Lyapunov
exponent $\lambda$, which is defined at the viscous scale. Hence, the
dynamics of a molecule is sensitive to the fluid motion at the viscous
scale whereas the velocity varies over a wide interval of
scales. Therefore the Lyapunov exponent varies in time and space over
scales from the inertial interval. We thus have an ``intermittent
picture'': In the regions where $\lambda<1/\tau$ one deals with a
Newtonian fluid with the viscosity $\nu$ of the solvent, whereas in
the regions where $\lambda>1/\tau$ the polymers are strongly stretched
and the effective viscosity can be much larger than $\nu$.

As the Reynolds number ${\rm Re}$ increases, the relative volume of
the regions with $\lambda>1/\tau$ increases and the averaged (over
space) viscosity grows. The average value of $R_{\rm op}$ also
grows. After it has reached the value of the order $R_{\rm max}$, the
back influence cannot grow anymore. It means that the effective
viscosity first grows and then decreases back to the solvent value
$\nu$. Note that the effective viscosity varies smoothly without
onset. As a consequence, the drag reduction also varies smoothly with
Re, having a maximum at some intermediate value. Experiments seem to
confirm our picture (see, e.g., \cite{95GB}).

To avoid misunderstanding let us stress that we consider conventional
turbulent flows which have the inertial interval of scales. In the
inertial interval the polymer back reaction is small compared to the
non-linear term in Eq. (\ref{navier}). In principle, in some region of
parameters the back reaction can be stronger than this non-linearity
everywhere and the properties of the fluid are drastically different
\cite{98GS}. This case requires a separate analysis.

We thank A.~Groisman for numerous talks that initiated this work. We
are indebted to G.~Falkovich for important remarks. Helpful
discussions with V.~Steinberg are greatfully acknowledged. We thank
M.~Chertkov for the possibility to read his work \cite{misha} before
its publication. The work was supported by the grants of Minerva
Foundation, by the Edward and Anna Mitchell Research Fund at the
Weizmann Institute and by the Landau-Weizmann Prize
program. E.B. acknowledges support by NEC Research Institute, Inc.

\end{multicols}
\end{document}